\documentclass[prl,aps,epsf,twocolumn,showpacs]{revtex4-1}
\usepackage[pdftex]{graphicx}
\usepackage{dcolumn}
\usepackage{bm}
\usepackage{epsfig}
\usepackage{latexsym}
\usepackage{amsmath}
\usepackage{color}
\usepackage{array}

\newcommand{\e}{\varepsilon}
\newcommand{\up}{\uparrow}
\newcommand{\down}{\downarrow}

\renewcommand{\(}{\left(}
\renewcommand{\)}{\right)}
\renewcommand{\[}{\left[}

\renewcommand{\v}[1]{\mathbf{#1}} % \v -> vector (bf)

\begin{document}
\title{Zero-bias peaks in spin-orbit coupled superconducting wires with and without Majorana end-states}
\author{Jie Liu$^1$}\thanks{These authors made equal contributions to this paper.}
\author{Andrew C. Potter$^2$}\thanks{These authors made equal contributions to this paper.}
\author{K.T. Law$^1$}\author{Patrick A. Lee$^2$}
\affiliation{$^1$Department of Physics, Hong Kong University of Science and Technology, Clear Water Bay, Hong Kong, China}
\affiliation{$^2$Massachusetts Institute of Technology 77 Massachusetts Ave. Cambridge, MA 02139}

\begin{abstract} One of the simplest proposed experimental probes of a Majorana bound-state is a quantized ($2e^2/h$) value of zero-bias tunneling conductance.  When temperature is somewhat larger than the intrinsic width of the Majorana peak, conductance is no longer quantized, but a zero-bias peak can remain.  Such a non-quantized zero-bias peak has been recently reported for semiconducting nanowires with proximity induced superconductivity.  In this paper we analyze the relation of the zero-bias peak to the presence of Majorana end-states, by simulating the tunneling conductance for multi-band wires with realistic amounts of disorder.  We show that this system generically exhibits a (non-quantized) zero-bias peak even when the wire is topologically trivial and does not possess Majorana end-states.  We make comparisons to recent experiments, and discuss the necessary requirements for confirming the existence of a Majorana state.
\end{abstract}
\maketitle

Recent proposals\cite{Fujimoto,Sato1,Sato2,AliceaPRB,LutchynWires,OregWires,SauSemiconductorSpinOrbit} to build topological superconductors from conventional spin-orbit coupled systems have sparked an active experimental effort to realize Majorana fermions and probe their predicted non-Abelian exchange statistics.    Tunneling from a normal wire into a topological superconducting wire with a Majorana end-state yields a quantized $G(0)=\frac{2e^2}{h}$ conductance peak at zero-bias\cite{Sengupta,LawResonantAR,Wimmer}.  This quantized zero-bias peak (ZBP) constitutes one of the simplest and most direct experimental probes for a Majorana fermion, and is likely to be the first test conducted on any putative topological superconducting wire.  The observation of quantized zero-bias conductance with $G(0)=\frac{2e^2}{h}$ ZBP requires temperature, $T$ to be sufficiently smaller than the intrinsic width, $\gamma$, of the Majorana peak, due to hybridization with the normal lead. For $T$ comparable to or somewhat larger than $\gamma$ a ZBP may still occur, but its no longer quantized and can take any value less than $\frac{2e^2}{h}$\cite{Sengupta,LawResonantAR,Wimmer}.  

A recent set of experiments on InSb nanowires coated with a superconducting NbTiN layer report the observation of non-quantized ZBP's when a magnetic field of sufficient strength is applied along the wire\cite{Mourik}. Similar results have since been reported by other groups\cite{Deng,Das}.  These experimental observations are qualitatively consistent with the existence of Majorana end-states, and constitute an important first step towards the realization of Majorana fermions in solid-state systems.  Given the potential significance of these findings, it is important to build a more quantitative understanding of the experimental system.  In particular, we would like to establish whether observed non-quantized ZBP's definitely correspond to thermally broadened peaks from Majorana end-states, or whether they could be produced by some other mechanism?  

\begin{figure}[ttt]
\begin{center}
\includegraphics[width = 3.4in]{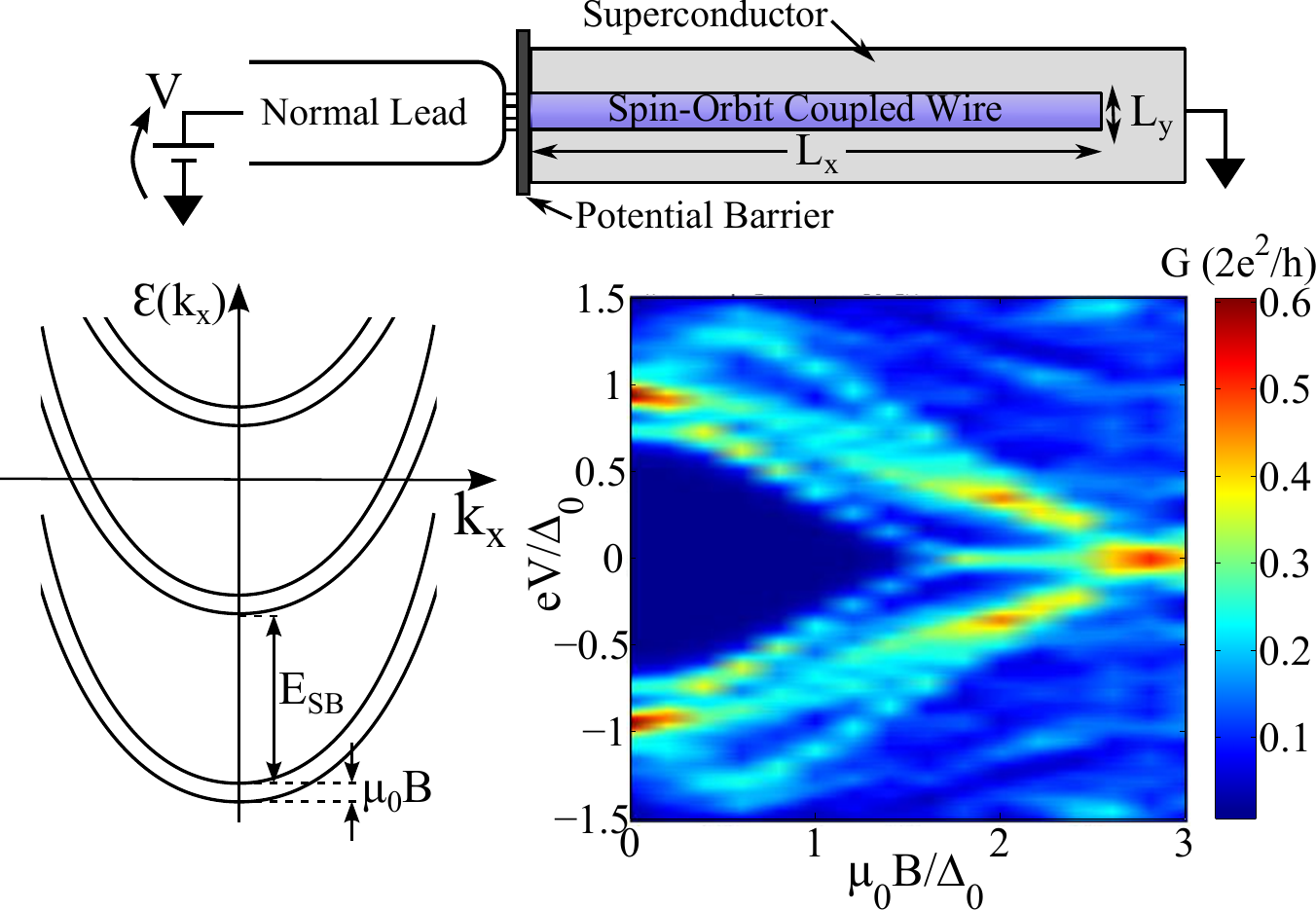}
\end{center}
\vspace{-.2in}
\caption{(Top) Schematic of tunneling geometry. (Lower Left) Dispersion of sub-bands in multi-band wire.  Each sub-band is split by $\mu_0B$ due to the magnetic field.  Majorana fermions appear only when an odd number of sub-bands is occupied.  (Lower Right) Color plot of tunneling conductance, $G$, at finite temperature as a function of applied field $\mu_0B$ and lead-wire voltage, $V$, for a multi-band wire with realistic amounts of disorder (see Fig.~\ref{fig:BSweepCurvesDirty} for detailed parameters).  A stable zero-bias peak appears despite the fact that there is no Majorana end-state.  At lower temperature, the peak is revealed to come from a cluster of low-energy states (see Fig.~\ref{fig:BSweepCurvesDirty}).}
\vspace{-.2in}
\label{fig:Schematic}
\end{figure}

To this end, we have conducted numerical simulations of tunneling conductance for spin-orbit coupled wires with proximity-induced superconductivity.  Our simulations use realistic energy scales appropriate for InSb wires, and consider the various experimentally relevant non-idealities including: multiple occupied sub-bands\cite{WimmerMultiBand,ACPMultiBand1,ACPMultiBand2,DasSarmaMultiband}, modest amounts of disorder\cite{ACPMultiBand1,ACPDisorder,StanescuDisorder,BrouwerDisorder}, and non-zero temperature.  Our study reveals important features absent in previous studies of clean- or very weakly disordered wires\cite{ChienHungSplitting,StanescuGapClosing,Pientka}.  

We find that, at non-zero temperature and in the presence of multiple sub-bands and weak disorder, zero-bias peaks generically occur even when the wire is in the topologically trivial regime and does not have Majorana end-states.  Furthermore, we find that the ZBP's persist even when disorder is sufficiently strong to destroy the topological phase and fuse the Majorana fermions on each side of the wire\cite{ACPMultiBand1,ACPDisorder,StanescuDisorder,BrouwerDisorder}.  Such ZBP's are also found outside the range in chemical potential where Majorana end states are expected in the clean limit, and are produced by ordinary fermion states localized to the wire ends and clustered near the Fermi-energy.  These states are in some sense, remnants of Majoranas, and appear and disappear with magnetic field in the same way as true Majorana end-states.  Therefore, we argue, that the only way to definitely rule out a non-topological origin to the ZBP is to lower temperature below the thermally broadened regime and observe a truly quantized zero-bias conductance peak, well isolated from other background states.

\noindent\textbf{Model --} 
We consider a three-dimensional rectangular wire of length $L_x$ along the $\hat{x}$ direction and cross sectional area $L_y\times L_z$ in the yz-plane.  The continuum Hamiltonian for the spin-orbit coupled wire without proximity induced superconductivity is:
\begin{align} H = &
\\
\sum_{\v{r}}&c^\dagger_{\v{r},\alpha}\(\frac{-\nabla^2}{2m}-\mu-i\alpha_R\hat{z}\cdot(\boldsymbol{\sigma}\times \nabla)-\mu_0\v{B}\cdot\boldsymbol{\sigma}\)_{\alpha\beta}c_{\v{r},\beta} \nonumber
%+\sum_{\v{r}}\Delta_0 c^\dagger_{\v{r},\up}c^\dagger_{\v{r},\down}+\text{h.c.}
\end{align}
Here $\alpha_R$ is the Rashba velocity, related to the spin orbit coupling by $E_\text{so} = \frac{1}{2}m\alpha_R^2$, $\mu_0=g\mu_B$ is the Zeeman coupling to the magnetic field $\v{B}$ taken throughout to point along the wire (in the $\hat{x}$ direction), and $\Delta_0$ is the proximity--induced pairing amplitude. 

To model this system, we approximate the continuum Hamiltonian by the following discrete tight binding Hamiltonian, defined on a $N_x\times N_y\times N_z$ site prism:  
\begin{align}
H_{\text{tb}} = &   \sum_{\v{r},\v{d}}c^\dagger_{\v{r}+\v{d},\alpha}\[-t\delta_{\alpha\beta}-iU_R\hat{z}\cdot\(\boldsymbol{\sigma}_{\alpha\beta}\times\hat{d}\)\]c_{\v{r},\beta}-
\nonumber\\
&-\sum_{\v{r}}c^\dagger_{\v{r},\alpha}\[\mu\delta_{\alpha\beta}+\mu_0\v{B}\cdot\boldsymbol{\sigma}_{\alpha\beta}\]c_{\v{r},\beta}+
%+\Delta_0\sum_{\v{r}}\(c^\dagger_{\v{r},\up}c^\dagger_{\v{r},\down}+\text{h.c.}\)
\nonumber\\
&+V_\text{imp}(\v{r})\sum_{\v{r}}c^\dagger_{\v{r},\alpha}c_{\v{r},\alpha}\end{align}
Here, we have included a random impurity potential $V_\text{imp}(\v{r})$, which is chosen independently for each site, identically distributed according to a Gaussian with variance $\overline{V(\v{r})V(\v{r}')}=W^2\delta_{\v{r},\v{r}'}$, where $\overline{(\cdots)}$ indicates averaging with respect to disorder configuration.

\begin{figure}[ttt]
\begin{center}
\includegraphics[width = 1.65in]{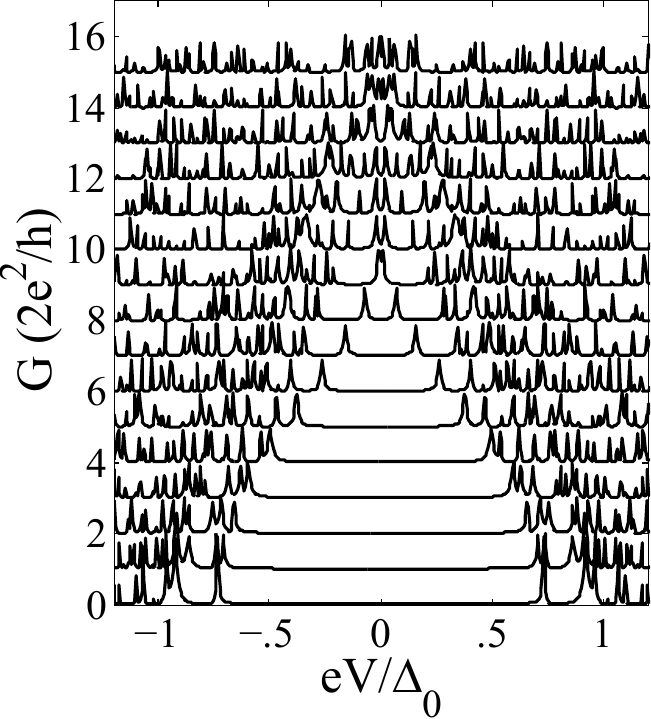} 
\includegraphics[width = 1.65in]{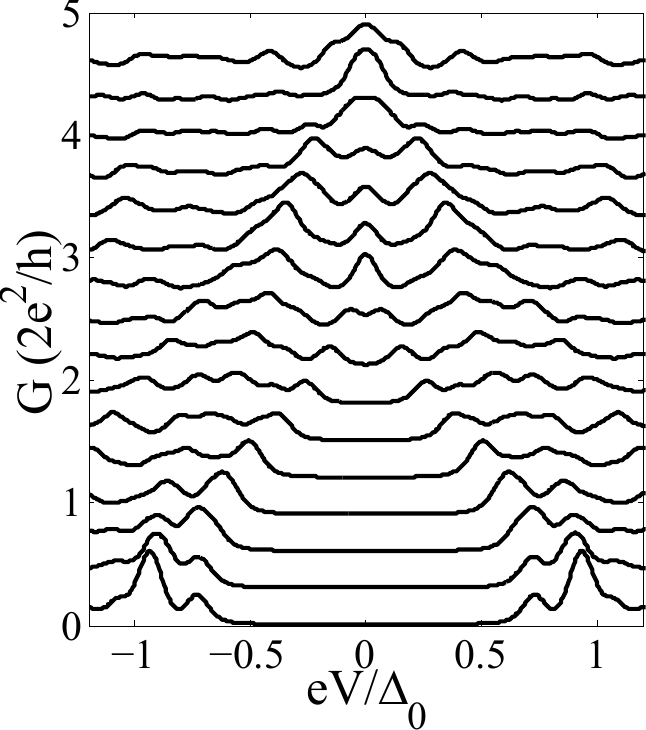}
\end{center}
\vspace{-.2in}
\caption{Conductance traces as a function of $\mu_0B$, corresponding to the color plot in Fig.~\ref{fig:Schematic}.  From bottom to top $\mu_0B$ ranges from 0 to $3\Delta_0$ in steps of $0.2\Delta_0$ (curves are offset for clarity).  $\mu=-172$ corresponding to cut $A$ in Fig.~\ref{fig:MuSweepColorDirty}, and to 6 occupied sub-bands (including spin).  ZBPs appear for $\mu_0B\approx \Delta_0$ just as for a Majorana end-state, despite having an even number of occupied sub-bands.  Wire dimensions are $N_y\times N_z\times N_x=6\times 5\times 180$. Tight-binding parameters: $W=12$, $t=36.5$, $U_R=2.7$, $\gamma_\text{SC}=2.5$, and $\gamma_\text{LW}=0.3$. }
\vspace{-.1in}
\label{fig:BSweepCurvesDirty}
\end{figure}

\begin{figure}[ttt]
\begin{center}
\hspace{-.1in}\includegraphics[width = 2.1in]{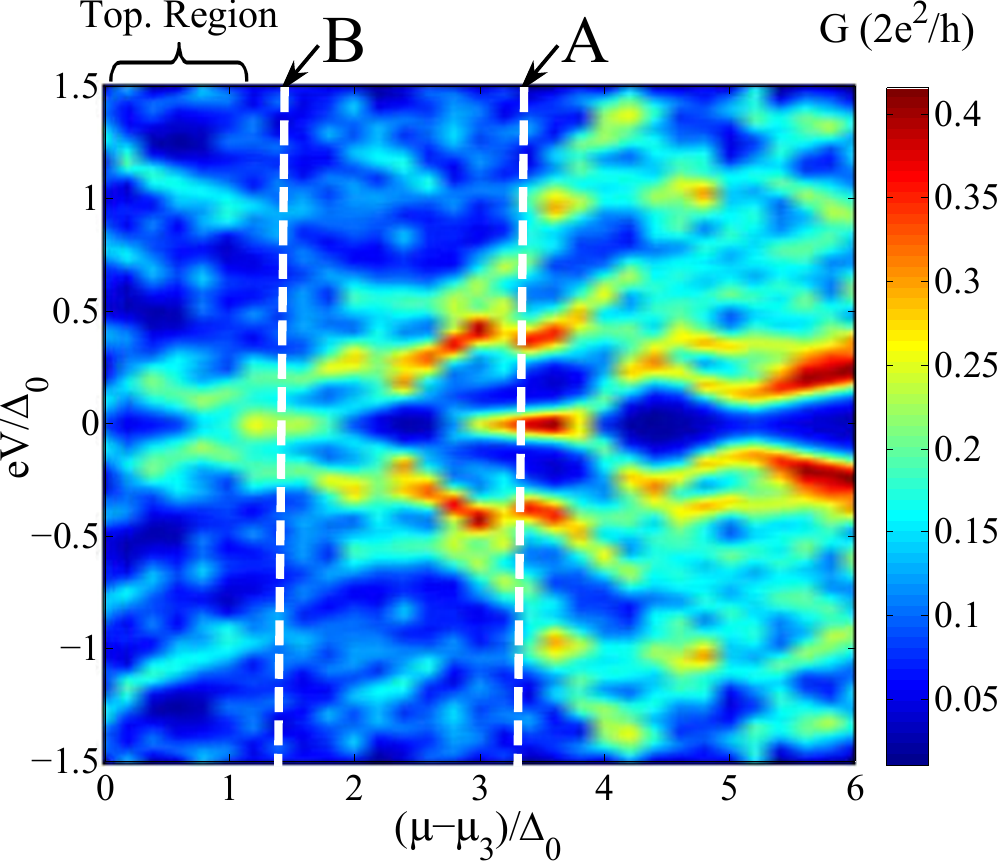}
\end{center}
\vspace{-.2in}
\caption{(Left) Color plot of conductance for same parameters as Fig.~\ref{fig:BSweepCurvesDirty}, but with fixed $\mu_0B=1.5\Delta_0$. $\mu$ is varied so that between 5-6 sub-bands are occupied, and is measured with respect to the center of the 3rd topological region located at $\mu_3=-175.2$.  Stable ZBPs occur for a wide-range of $\mu$, but have no topological origin.  Detailed $\mu_0B$ dependence is shown for $\mu$ corresponding to the dashed line A (see Figs.~\ref{fig:Schematic} and \ref{fig:BSweepCurvesDirty}) and line B (see Fig.~\ref{fig:BRot}).  The topological region in the absence of disorder, indicated by ``Top. Region", does not exhibit ZBPs for this disorder configuration.}
\label{fig:MuSweepColorDirty}
\vspace{-.1in}
\end{figure}

Table \ref{tbl:tbparms} relates the tight-binding parameters to the continuum model and gives estimated values for InSb nanowires.  There is considerable uncertainty in the estimated spin-orbit strength, which was measured without the superconducting layer\cite{Nadj-PergeSpinOrbitDot}.  Since $E_\text{so}$ derives solely from the inversion symmetry breaking potential of the substrate-wire and superconductor-wire interfaces, the actual value could be rather different than $50\mu$eV, and one should consider the possibility that $E_\text{so}$ is much smaller (or larger). The mean-free path from disorder is $\ell = v_F\tau$ where $\tau^{-1} \approx 2\pi \frac{W^2a}{L_yL_z}N(\mu)$ is the elastic scattering rate due to impurities.  Here $N(\mu)$ is the 1D density of states at the chemical potential: $N(\mu) = \sum_n\frac{1}{2\pi a\sqrt{t(\mu-\e_n)}}$, where the sum is over occupied sub-bands labeled by $n$ and having band-bottoms located at energy $\e_n$.  Transport experiments estimate $\ell\approx 3\mu$m (again without a superconducting layer)\cite{PlissardMFP}. Since these measurements were done at large source-drain bias, this value reflects a sort of average over the lowest 3-4 sub-bands, and should be taken as a rough guide. 
\begin{center}
\begin{table}[tb]
\begin{tabular}{c c c c}
\hline\hline
Parameter & Symbol & TB Equivalent & InSb Value\\
\hline
Wire Diameter& $L_{y,z}$ & $N_{y,z}a$ & $100$nm
\\
Wire Length & $L_x$ & $N_xa$ & $\approx 2$-$3\mu$m
\\ 
Band Mass &m & $(2ta^2)^{-1}$& 0.015$m_e$ 
\\
Spin-Orbit & $E_\text{so} = \frac{1}{2}m\alpha_R^2$ & $U_R=\sqrt{E_\text{so}t}$& 50$\mu$eV 
\\
Induced SC Gap& $\Delta_0$ & $\Delta_0$ & 250$\mu$eV\cite{Nadj-PergeSpinOrbitDot}
\\
Mean Free Path& $\ell$ & (see text) & $\approx 3\mu$m\cite{PlissardMFP}
\\
Min. Temperature & T & & $0.03\Delta_0$
\\\hline
\end{tabular}
\caption{Tight-binding (TB) model parameters, and estimated values for InSb/NbTiN experiment\cite{Mourik}. $a$ denotes lattice spacing in the TB model.}
\label{tbl:tbparms}
\end{table}
\vspace{-.3in}
\end{center}

Since only the outer-boundary of the wire is in contact with the superconductor, there will in general be different proximity induced gaps for different sub-bands.  These multi-band effects can be important for reproducing the observed data for InSb wires.  There, coherence peaks are observed at energy, $\Delta_0\approx 250\mu$eV, but non-zero conductance occurs within the proximity induced ``gap".  The shape of this sub-gap conductance is not consistent with a Lorenzian broadening of the coherence peaks from coupling to the lead, but could be explained with a multi-gap scenario in which some sub-bands have proximity induced gaps smaller than $\Delta_0$.  However, we have so far not been able to reproduce, in detail, the large sub-gap tunneling conductance at $\mu_0B=0$ observed in the experiment.

\noindent{\bf Tunneling Conductance -} We use the iterative transfer matrix method to construct the Green's function for the end of the wire, and compute the scattering matrix from the Green's function\cite{LeeFisher,Wimmer}. Experimentally, the wires are terminated by a large gap superconductor and only Andreev reflection contributes to tunneling current:
\begin{align} I(V) = &\frac{2e^2}{h}\int d\e \(f(\e-eV)-f(\e)\)\text{tr}|\hat{r}_{eh}(\e)|^2\end{align}
where $\hat{r}_{eh}(\e)_{ij}$ is the electron-hole part of the reflection matrix from channel $i$ to channel $j$ in the lead, and $f$ is the Fermi distribution.  Throughout, we take the lead to have position and energy independent density of states $N_0 = 1/(\pi v_F)$, and model the tunneling barrier by a weak hopping link with hopping strength $t_\text{LW}$.  The lead-wire coupling is characterized by $\gamma_\text{LW}=N_0|t_\text{LW}|^2$.  

Proximity induced superconductivity is modeled by coupling the boundary of each cross-section in the yz-plane to an infinite superconductor, producing the self-energy\cite{SauProximity,ACPDisorder}:
\begin{equation} \Sigma_\text{SC}(\omega_+,y,z) = \hat{\mathcal{P}}_{\text{edge}}\frac{\gamma_\text{SC}\(\omega_+-\Delta_B\tau_3\)}{\sqrt{\Delta_B^2-\omega_+^2}}
\end{equation}
The projection $\hat{\mathcal{P}}_{\text{edge}}(y,z)=1$ if $(y,z)$ is on the outer edge of the wire ($y=0$) and is zero otherwise. Here, $\omega_+=\omega+i\eta$ where $\eta$ positive and infinitesimal, $\tau_3$ is the z-Pauli matrix in the Nambu/Gorkov particle-hole basis:
$\Psi(\v{k}) = \begin{pmatrix} c_{\v{k},\up} & c_{\v{k},\down} & -c_{-\v{k},\down}^\dagger & c_{\v{k},\up}^\dagger \end{pmatrix}^T$, where $c_{k,s}$ destroys an electron with momentum $\v{k}$ and spin $s\in\{\up,\down\}$.
$\Delta_B$ is the pairing gap of the adjacent bulk superconductor $\gamma_\text{SC} = \pi N_B(0)|\Gamma|^2$ is the strength of coupling between the wire and superconductor, $N_B(0)$ is the density of states of the bulk superconductor near the Fermi-surface, and $\Gamma$ is the wire-superconductor tunneling amplitude.

\begin{figure}[ttt]
%\begin{center}
%\includegraphics[width = 1.65in,height=2.25in]{CleanBSweepL180.pdf}
%\includegraphics[width = 1.65in,height=2.25in]{CleanBSweepL180T.pdf}
%\end{center}
\begin{center}
\includegraphics[width=1.5in]{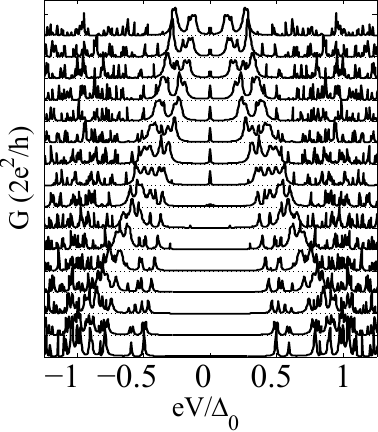}
\includegraphics[width=1.5in]{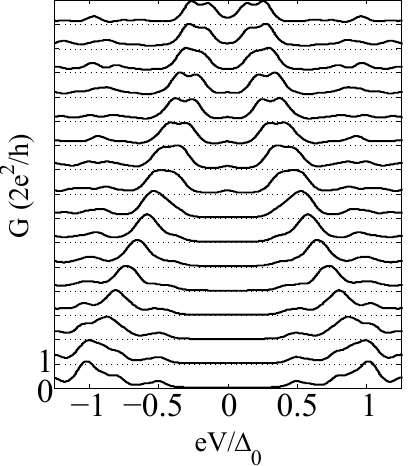}
\includegraphics[width = 2in]{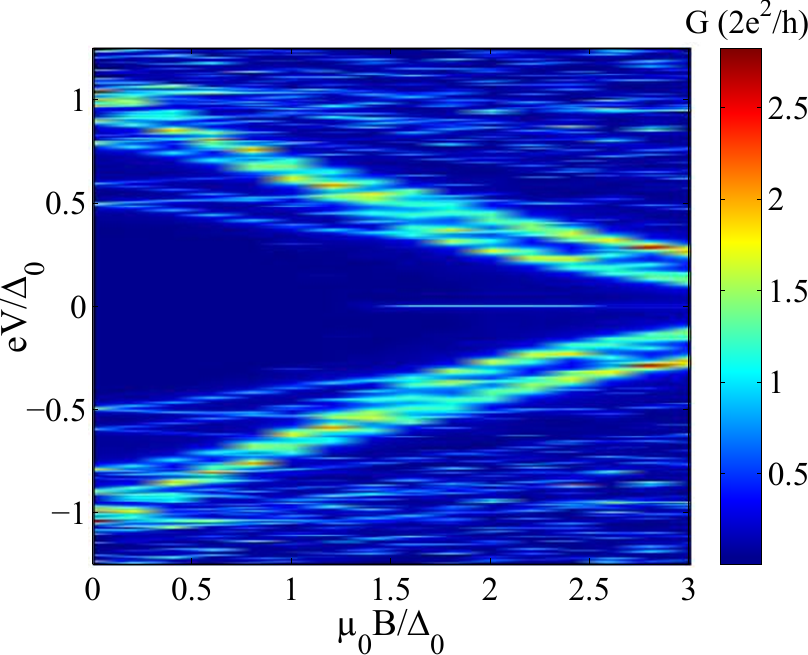}
\end{center}
\vspace{-.2in}
\caption{Signature of an ideal Majorana state for a long, clean wire ($N_x=360$, $W=0$).  (Top Row) $\mu_0B$ dependence for $\mu$ in the center of the 6th topological region (11 occupied sub-bands), at $T=0$ (left) and $T=0.03\Delta_0$ (right).  (Bottom) Corresponding color plot for $T=0$. }
\label{fig:BSweepClean}
\vspace{-.1in}
\end{figure}

\noindent\textbf{Conditions for Majorana End-states --}
In a multiband wire, when $\mu_0B>\Delta_0$, one can think of each sub-band as contributing a Majorana end-state which then mix.  For an even number of occupied sub-bands, the Majoranas fuse into ordinary fermions and are pushed away from zero-energy.  By contrast, for an odd number of occupied sub-bands a single Majorana state always remains at zero-energy\cite{ACPMultiBand1}. More quantitatively, to observe a Majorana, $\mu$ must fall within specific intervals of size $\approx \pm|\mu_0B-\Delta_0|$.  In \cite{Mourik} ZBP's are observed for $\mu_0B\gtrsim \Delta_0\approx 250\mu$eV.  Indicating that Majorana end-states can exist only in narrow regions of $\mu$ of size $\approx 250\mu$eV, or $\approx \frac{1}{10}$ of the typical sub-band spacing $E_\text{sb}\approx 2.5$meV. 
% Since the value of $\mu$ is unknown a priori, without tuning chemical potential over a large range one would have to rely on good fortune to produce samples that intrinsically fall close to this narrow window.\cite{VanHovePinningEndNote}  

Problems can arise for short wires\cite{Kitaev,ChienHungSplitting}.  If the wire is not sufficiently long, then the Majorana states on each end of the wire can overlap, splitting into ordinary fermion states.  For example, the InSb wires in \cite{Mourik}, had aspect ratios of $L_x\approx 20-30L_y$.  Tight-binding simulations of such a wire, with the parameters as in Table~\ref{tbl:tbparms}, show that Majorana end-states hybridize with splitting of a few percent of $\Delta_0$ even for clean wires.  This issue is exacerbated by disorder, which is strongly pair-breaking\cite{ACPDisorder,BrouwerDisorder,StanescuDisorder} and reduces the gap protecting Majorana end-states, allowing them to spread out and hybridize more strongly.  Indeed, realistic amounts of disorder corresponding to the estimated $\ell$ readily destroy the quantized conductance peak from Majorana end-states.

Despite these issues, ZBP's are still observed in InSb wires. This suggests that ZBP might occur rather generally and may not necessarily tied to the presence of zero-energy Majorana end-states.  Below, we will show that this is indeed the case, and that stable ZBP's frequently appear without Majorana states.

\noindent{\bf Tunneling Data-} To start, we first consider an idealized case, showing quantized Majorana peaks. Fig.~\ref{fig:BSweepClean} shows conductance data for a long, perfectly clean wire of length $N_x=360$  in the center of the 6th topological region (i.e. with 11 occupied sub-bands including spin).  The left panel shows traces at $T=0$.  As $\mu_0B$ is increased the bulk pairing gap collapses, leaving behind a single quantized Majorana peak.  At finite temperature, $T=0.03\Delta_0$, the narrow Majorana peak is thermally broadened and greatly reduced in height. Fig.~\ref{fig:BSweepClean} along with a color plot of the $T=0$ curves.  We note that, even for a clean wire, the observation of this quantized peak requires much 2-3 times longer wires than those used in \cite{Mourik}, and an order of magnitude larger than in \cite{Das}.

\begin{figure}[ttt]
\begin{center}
\includegraphics[width = 1.2in,height=1.55in]{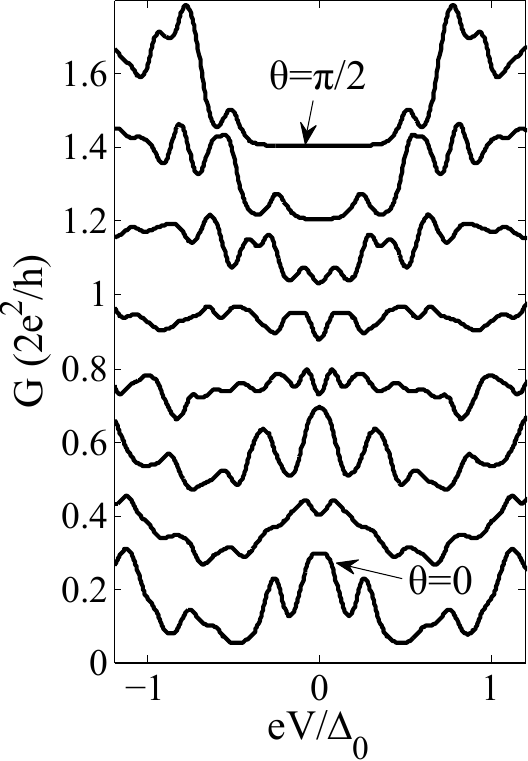}\hspace{.1in}
\includegraphics[width=2in]{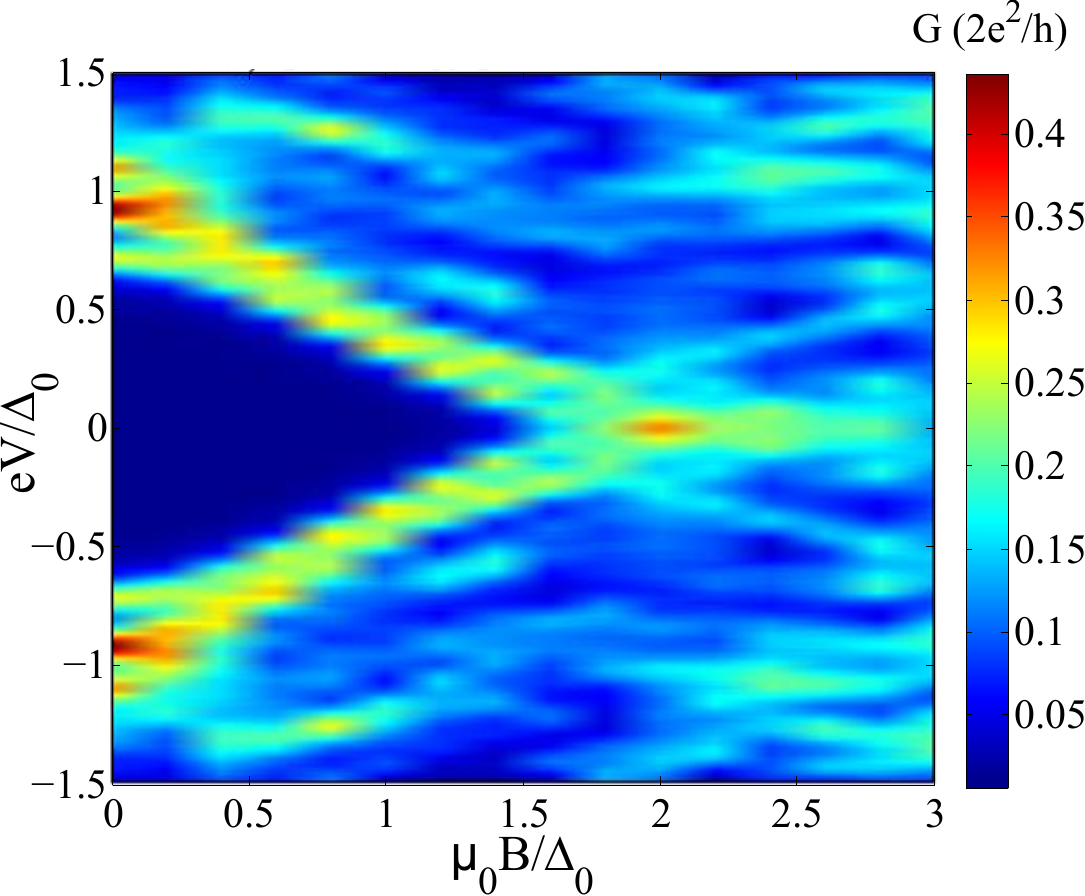}
\end{center}
\vspace{-.2in}
\caption{(Left) Angle dependence of non-topological ZBP is similar to that of a true Majorana derived ZBP (curves are offset for clarity). $\theta$ measures angle of $\v{B}$ and the x-axis in the x-y plane. (Right) Color plot of $\mu_0B$ dependence for $\mu$ corresponding to cut B in Fig.~\ref{fig:MuSweepColorDirty}.}
\vspace{-.2in}
\label{fig:BRot}
\end{figure}

Under realistic experimental conditions for semiconducting wires, disorder is expected to play an important role\cite{ACPDisorder,BrouwerDisorder,StanescuDisorder}. The pair-breaking effects of disorder in a magnetic-field tend to collapse the proximity induced pairing gap, causing Majorana peaks to split away from zero-energy. Furthermore, disorder reduces the mini-gap to end-states from other occupied sub-bands.  In the absence of disorder, these end-states are pushed up above induced gap where they get absorbed by a continuum of extended states.  In a field, disorder reduces the mini-gap splitting, causing these end-states to cluster near zero-energy.  Our simulations show that, that clusters emerging from mini-gap states frequently remain localized near the end of the wire, and at temperature larger than the mini-gap splitting merge into a single ZBP.  

Fig.~\ref{fig:MuSweepColorDirty} shows a color plot of tunneling conductance for fixed $\mu_0B=1.5\Delta_0$ and $T=0.03\Delta_0$, as a function of $\mu$ near the 3rd topological region (with 5-6 sub-bands occupied).  The modest amount of disorder included here ($W=12$) corresponds to a very long mean-free path ($\ell\approx 10\mu$m), but is nevertheless sufficient to destroy Majorana end-states.  Despite this, zero-bias peaks frequently emerge (we note that these peaks appear even more readily for stronger disorder, for example with $W$ corresponding to $\ell\approx 2.5\mu$m relevant to InSb wires).  These peaks are typically stable over intervals of $\Delta\mu\approx \Delta_0$ although they are no-longer tied to the topological region $|\mu-\mu_3|<\sqrt{(\mu_0B)^2-\Delta_0^2}$.  
Fig.~\ref{fig:BSweepCurvesDirty} shows a typical example of this non-topological ZBP developing as a function of applied field, corresponding to $\mu$ along cut $A$ in Fig.~\ref{fig:MuSweepColorDirty}.  At $T=0$ (left), one can resolve the ZBP into multiple peaks near zero-energy, however at $T=0.03\Delta_0$ (right) the peaks are smeared into a single ZBP.  

These peaks appear and disappear under very similar magnetic field conditions as true Majorana states would: 1) they appear only when $\mu_0B\gtrsim \Delta_0$ when pair-breaking effects or disorder become important, 2) they disappear when the field is rotated to point along the wire (see Fig.~\ref{fig:BRot}).  However, unlike ZBP's tied to Majorana states, these non-topological ZBP's appear commonly throughout the range of chemical potentials between adjacent sets of sub-bands, meaning that they would be more readily observed without strongly tuning $\mu$.  To illustrate this last point, Fig.~\ref{fig:BRot} shows a color plot of the B-field dependence of conductance for $\mu$ corresponding to slice B in Fig.~\ref{fig:MuSweepColorDirty}.  The results are similar to slice A, and further simulations show that this B-field dependence is generic whenever a ZBP is stable over an interval of $\mu$.

The features described here depend strongly on the size of the spin-orbit coupling.  For weaker spin-orbit coupling, non-topological ZBPs appear even more readily and at even weaker levels of disorder.  Conversely, if $E_\text{so}$ is substantially larger than estimated in \cite{Nadj-PergeSpinOrbitDot}, it becomes difficult to reconcile the observations with a ZBP of non-topological origin.

\noindent\textbf{Discussion -- }  We have shown that, under experimentally realistic conditions for semiconducting wires with rather modest amounts of disorder, Majorana end-states are destroyed and do not give rise to quantized ZBPs.  Nevertheless, at finite temperature, ZBPs of a non-topological origin often appear due to the pair-breaking effects of disorder, which lead to clusters of low-energy states localized near the wire end.  These non-topological ZBPs are typically stable with respect to variations of chemical potential and magnetic field, and appear and disappear under nearly identical conditions to those of true Majorana peaks.  

These results strike a note of caution for interpreting recent experimental evidence of Majorana states in tunneling data\cite{Mourik,Deng,Das}.  In order to truly identify Majorana end-states and rule out a non-topological origin of observed ZBPs, it is crucial to push to lower temperatures, and observe a quantized conductance peak. At sufficiently low temperature non-topological ZBPs will reveal themselves as clusters of states, and can be distinguished.  Spurious disorder induced peaks can also be ruled out by more complicated measurements, such as the $4\pi$ Josephson effect\cite{Kitaev}, which does not survive once Majoranas are destroyed by disorder\cite{LawFJE}.

Our simulations indicate that to realize Majorana states in semiconductors, one likely needs to produce substantially longer and cleaner wires.  These difficulties suggest that it may be beneficial to seek materials with larger spin-orbit\cite{ACPMetalFilm}.

\noindent\textbf{Acknowledgements -- } ACP and PAL acknowledge support from DOE Grant No. DEFG0203ER46076.  JL and KTL are supported by HKRGC through HKUST3/CRF09 and DAG12SC01.

During the completion of this manuscript a related work appeared by Bagrets and Altland, which finds a disorder induced zero-bias peak in the disorder-averaged density of states that may be related to the ZBP discussed here.  However, we caution that since the tunneling conductance into one point of the wire is not a self-averaging property, it may be problematic to make comparisons from disorder averaged properties.


\begin{thebibliography}{99}
% Original Proposals
\bibitem{Fujimoto}
S. Fujimoto, Phys. Rev. B. {\bf 77}, 220501(R) (2008)
\bibitem{Sato1}
M. Sato, Y. Takahashi, and S. Fujimoto, Phys. Rev. Lett. {\bf 104}, 040502 (2010)
\bibitem{SauSemiconductorSpinOrbit}
J.D. Sau, R.M. Lutchyn, S. Tewari, and S. Das Sarma, Phys. Rev. Lett. {\bf 104}, 040502 (2010)
\bibitem{Sato2}
M. Sato, Y. Takahashi, and S. Fujimoto, Phys. Rev. B {\bf 82}, 134521 (2010)
\bibitem{AliceaPRB}
J. Alicea Phys. Rev. B {\bf 81}, 125318 (2010)
\bibitem{LutchynWires}
R.M. Lutchyn, J.D. Sau, and S. Das Sarma, Phys. Rev. Lett. {\bf 105}, 077001 (2010)
\bibitem{OregWires}
Y. Oreg, G. Refael, and F. von Oppen, Phys. Rev. Lett. {\bf 105}, 177002 (2010)
% Previous Work on Tunneling Conductance
\bibitem{Sengupta}
K. Sengupta, Igor Žutic, Hyok-Jon Kwon, Victor M. Yakovenko, and S. Das Sarma, Phys. Rev. B {\bf 63}, 144531 (2001)
\bibitem{LawResonantAR}
K.T. Law, P.A. Lee, and T.K. Ng  Phys. Rev. Lett. {\bf 103}, 237001 (2009) 
\bibitem{Wimmer}
A.R. Akhmerov, J.P. Dahlhaus, F. Hassler, M. Wimmer, C.W.J. Beenakker, Phys.Rev.Lett. {\bf 106}, 057001 (2011)
% Nanowires Experiments
\bibitem{Mourik}
V. Mourik, K. Zuo, S.M. Frolov, S.R. Plissard, E.P.A.M. Bakkers, and L.P. Kouwenhoven, Science {\bf 336}, 1003 (2012)
\bibitem{Deng}
M. T. Deng, C.L. Yu, G.Y. Huang, M. Larsson, P. Caroff, H.Q. Xu, arXiv:1204.4130 (2012)
\bibitem{Das}
A. Das, Y. Ronen, Y. Most, Y. Oreg, M. Heiblum, H. Shtrikman, arXiv:1205.7073 (2012)
% Multi-band
\bibitem{WimmerMultiBand}
M. Wimmer, A. R. Akhmerov, M. V. Medvedyeva, J. Tworzydlo, C.W.J. Beenakker, Phys. Rev. Lett. {\bf 105} 046803, (2010).
\bibitem{ACPMultiBand1}
A.C. Potter and P.A. Lee, Phys. Rev. Lett. {\bf 105}, 227003
(2010).
\bibitem{ACPMultiBand2} 
A.C. Potter and P.A. Lee, Phys. Rev. B {\bf 83}, 094525 (2011).
\bibitem{DasSarmaMultiband}
R.M. Lutchyn, T.D. Stanescu, and S. Das Sarma, Phys. Rev. Lett. {\bf 106} 127001, (2011).
% Disorder
\bibitem{ACPDisorder}
A.C. Potter and P.A. Lee, Phys. Rev. B {\bf 83}, 184520 (2011) 
\bibitem{BrouwerDisorder}
P.W. Brouwer, M. Duckheim, A. Romito, and F. von Oppen, arXiv:1104.1531v1 (2011); P.W. Brouwer, M. Duckheim, A. Romito, and F. von Oppen, arXiv:1103.2746v1 (2011).
\bibitem{StanescuDisorder}
T.D. Stanescu, R.M. Lutchyn, and S. Das Sarma, Phys. Rev. B {\bf 84}, 144522 (2011)
% Finite-Size Splitting
\bibitem{ChienHungSplitting}
C.-H. Lin, J.D. Sau, S. Das Sarma, arXiv:1204.3085 (2012)
% Clean Conductance Studies
\bibitem{StanescuGapClosing}
T.D. Stanescu, S. Tewari, J.D. Sau, S. Das Sarma  arXiv:1206.0013, (2012)
\bibitem{Pientka}
F. Pientka, G. Kells, A. Romito, P.W. Brouwer, F. von Oppen,  arXiv:1206.0723 (2012).
% Characterization of InSb Wires
\bibitem{Nadj-PergeSpinOrbitDot}
S. Nadj-Perge, V.S. Pribiag, J.W.G. van den Berg, K. Zuo, S.R. Plissard, E.P.A.M. Bakkers, S.M. Frolov, and L.P. Kouwenhoven
Phys. Rev. Lett. {\bf 108}, 166801 (2012)
\bibitem{PlissardMFP}
S. R. Plissard et al., Nano Lett. {\bf 12}, 1794 (2012).
% Conductance Transmission
\bibitem{LeeFisher}
P.A. Lee and D.S. Fisher, Phys. Rev. Lett. {\bf 47}, 882 (1981); D.S. Fisher and P.A. Lee Phys. Rev. B {\bf 23}, 6851 (1981).
% Proximity Effect
\bibitem{SauProximity}
J.D. Sau, R.M. Lutchyn, S. Tewari, S. Das Sarma,  Phys. Rev. B {\bf 82}, 094522 (2010)
%
\bibitem{Kitaev}
A. Kitaev, arXiv:cond-mat/0010440 (2000).
% notes
%\bibitem{VanHovePinningEndNote}
%Due to square-root Van-Hove singularity in the 1D wire dispersion, there is some weak pinning of the chemical potential near the sub-band bottoms.  In general, this pinning effect is rather weak, (typically, it actually reduces the fraction of densities over which Majorana states appear).
%
\bibitem{LawFJE}
K.T. Law and P.A. Lee, Phys. Rev. B {\bf 84}, 081304 (2011).
\bibitem{ACPMetalFilm}
A.C. Potter and P.A. Lee, Phys. Rev. B {\bf 85}, 094516 (2012)
%
\bibitem{Bagrets}
D. Bagrets and A. Altland, arXiv:1206.0434 (2012)

\end{thebibliography}
\end{document}